# Anomalous phonon dispersion of CaFe$_2$As$_2$ explored by inelastic neutron scattering

R. Mittal[1,2], L. Pintschovius[3], D. Lamago[3,4], R. Heid[3], K-P. Bohnen[3], D. Reznik[3], S. L. Chaplot[2], Y. Su[1], N. Kumar[5], S. K. Dhar[5], A. Thamizhavel[5], and Th. Brueckel[1,6]

[1]*Juelich Centre for Neutron Science, IFF, Forschungszentrum Juelich, Outstation at FRM II, Lichtenbergstr. 1, D-85747 Garching, Germany*

[2]*Solid State Physics Division, Bhabha Atomic Research Centre, Trombay, Mumbai 400 085, India*

[3]*Forschungszentrum Karlsruhe, Institut für Festkörperphysik, P.O.B. 3640, D-76021 Karlsruhe, Germany*

[4]*Laboratoire Leon Brillouin, CEA-Saclay, F-91191 Gif sur Yvette Cedex, France*

[5]*Department of Condensed Matter Physics and Material Sciences, Tata Institute of Fundamental Research, Homi Bhabha Road, Colaba, Mumbai 400 005, India*

[6]*Institut fuer Festkoerperforschung, Forschungszentrum Juelich, D-52425 Juelich, Germany*

We measured phonon dispersions of CaFe$_2$As$_2$ using inelastic neutron scattering and compared our results to predictions of density functional theory (DFT) in the local density approximation (LDA). The calculation gives correct frequencies of most phonons if the experimental crystal structure is used, except observed linewidths/frequencies of certain modes were larger/softer than predicted. Strong temperature dependence of some phonons near the structural phase transition near 172 K, may indicate strong electron phonon coupling and/or anharmonicity, which may be important for superconductivity.

**PACS numbers:** 74.25.Kc, 78.70.Nx, 63.20.-e, 71.38.-k

**Keywords**: Superconductivity, Neutron inelastic scattering, Phonons in crystal lattice, Electron-phonon coupling

The discovery of superconductivity at temperatures exceeding 50 K in iron arsenide compounds with general compositions RFeAsO (R = rare earth) and $MFe_2As_2$ (M = alkaline earth metal) and MFeAsF has attracted great interest [1-15] in these materials. At present, it is hotly debated whether these compounds are unconventional metals similar to the cuprate superconductors or can be understood within the same theoretical framework as conventional intermetallic compounds like the borocarbides or $MgB_2$. The parent compounds show [1-4] a crystallographic transition from a tetragonal to an orthorhombic phase, accompanied by antiferromagnetic spin ordering between 110 K to 200 K. Superconductivity appears either at a critical doping level of the parent compound, or by application of pressure [5,6] above a critical value. However, it remains unclear whether doping is essential to produce an optimal electron or hole concentration or simply to suppress the structural and magnetic instabilities. The fact that superconductivity can be induced as well by applying pressure – which suppresses the structural phase transformation and the magnetic ordering - points to the latter. The role of the phonons for the mechanism of superconductivity is not known at present.

DFT calculations predict weak electron-phonon coupling [9] with a negligible contribution to the superconductivity mechanism. Experimental information on the phonon properties of these compounds is, however, scarce, mainly because of the lack of suitable single crystals. Inelastic x-ray scattering investigation [10] of the phonon density of states in $LaFeAsO_{1-x}F_x$ and NdFeAsO, as well as measurements of a few phonon branches [10(a),11] on single crystals of $BaFe_2As_2$ and $PrFeAsO_{1-y}$ showed that DFT is only moderately [10(a),12] successful in predicting phonon frequencies in these compounds. The phonon density of states in $BaFe_2As_2$, $Sr_{0.6}K_{0.4}Fe_2As_2$ and $Ca_{0.6}Na_{0.4}Fe_2As_2$ was investigated on polycrystalline samples using inelastic neutron scattering [13,14]. Empirical models used to analyze the data again had limited success. After having grown a relatively large single crystal of $CaFe_2As_2$ (15 mm × 10 mm ×0.4 mm), we embarked on an inelastic neutron scattering study of the phonon dispersion to find out whether or not the phonon properties are really anomalous and whether there are indications of a strong electron-phonon coupling. In parallel, nonmagnetic DFT calculations were carried out for the same compound, using both the experimental and optimized structure. We will show that DFT correctly predicts most phonon frequencies for the experimental structure with the notable exception of certain phonons. To investigate the effect of magnetism, also spin-polarized DFT calculations for the optimized orthorhombic structure with the observed antiferromagnetic order were performed.

Single crystals of $CaFe_2As_2$ (15 mm × 10 mm ×0.4 mm) were grown from a high temperature solution using Sn as flux [15]. The details of crystal characterization are given in Ref. [15]. The neutron measurements were performed on the 1T1 triple-axis spectrometer at the Laboratoire Léon Brillouin, Saclay. Measurements were done with pyrolithic graphite (PG002) as a monochromator and analyzer. Most measurements were carried out at 300K with open collimations and double focusing on both the analyzer the monochrometor. Selected phonons were studied as a function of temperature down to T = 100 K, which is well below the magnetic/structural phase transition at 172K.

The calculations were carried out within the framework of the LDA and GGA using a mixed basis pseudopotential method [16]. A density functional perturbation approach was used for calculating the phonon frequencies and phonon eigenvectors [17]. We employed norm-conserving pseudopotentials and a plane-wave cutoff of 22 Ryd, augmented by local functions at the Ca and Fe sites. Brillouin zone (BZ) summations were done with a Gaussian broadening technique using a broadening of 0.2 eV and 40 wavevector points in the irreducible part of the BZ. For the non-magnetic LDA calculations, we started by minimizing the total energy as a function of the lattice parameters a and c and the internal parameter characterizing the position of the As atoms along c. The optimized structure was subsequently used for the phonon dispersion calculation. However, as reported previously [18,19], we soon realized that the optimized structure is relatively far away from the experimental one. In particular, the c-axis came out much shorter than observed. The agreement between the calculated and experimental phonon frequencies was also poor. In particular, the transverse acoustic phonons propagating along c are stiffer by a factor of two in the calculation compared to experiment (i.e. the corresponding force constant is four times larger!) Calculations using the experimental lattice constants but an optimized internal parameter have shown that optimization of the internal parameter alone already produces serious discrepancies between theory and experiment.

At first glance, the fact that the optimized structure is far away from the experimental one might be considered a blatant failure of DFT-LDA. However $CaFe_2As_2$ undergoes a first order transition to a "collapsed" structure at a pressure as low as 3.5 kbar [6] and table 1 shows that the calculated optimized structure is very close to that of the high pressure phase. The ambient pressure and 3.5kbar crystal structures should have nearly the same free energy, but LDA predicts a much lower energy for the high pressure phase even at ambient pressure [19]. One might expect that the experimental structure corresponds to a local minimum of the energy versus structure surface, but there is no evidence in the non-polarized calculations for a second minimum. It has been proposed that a large unordered Fe moment, which would have a strong effect on the structure is always present in these systems [20,18] and we performed spin-polarized calculations as well. LDA did not reveal any evidence for a second energy minimum either, and still predicted the nonmagnetic collapsed phase. In contrast, recent magnetic GGA calculations gave a structure close to the ambient temperature structure with a very large moment showing that the optimized structure strongly depends on the approximation for the exchange-correlation functional. References 18 and 19 reported similar results.

The internal structural parameter, the As-z position, controls the Fe-As bond length and has a large influence on the related force constants. Theoretical optimization gives values significantly closer to the experimental ones if magnetism is taken into account [18,19]. To mimic this effect, we performed further phonon calculations using the experimental structure as input. For our non-magnetic calculation this comes at the price that the structure is not force-free. Still such a calculation works because the force constants obtained in linear response calculations are related to second derivatives of the total energy with respect to atomic displacements, while atomic forces are given by first derivatives. Thus it is possible to have a non-relaxed structure with well behaving phonons. A subtle consequence is the violation of the rotational invariance, which, however, manifests itself mainly in the slopes of the acoustic branches. Figure 1 shows that DFT in the LDA is quite successful in predicting the phonon frequencies in $CaFe_2As_2$ if, instead of the relaxed structure, the experimental one is used. It is important to emphasize here that one must impose a nearly perfect tetrahedral environment of the Fe atoms as observed in experiment in order to obtain best agreement between calculated and experimental phonon frequencies. However, even in this case it is worse than in many other compounds, including conventional superconductors with high $T_c$ like $MgB_2$ [21]. This finding is similar to the previous observations on the Ba122 compounds by inelastic x-ray scattering [10(a)].

Although the calculations based on the experimental structure appear to be more accurate, some important differences with experiment remain. The main one is for the phonons of $\Delta_3$ symmetry between q=(0.5 0 0) and (1 0 0). Some appear around 19 meV but predicted at 22meV and others are observed around 16 meV but predicted 2 meV lower. One must also keep in mind that the good agreement for other phonons may be somewhat misleading in the sense that the predicted eigenvectors may differ from the experimental ones even where the phonon frequencies agree. In experiment, phonon eigenvectors determine observed phonon intensities. When phonons are nearly degenerate as in $CaFe_2As_2$ near 20 meV, different phonon branches may hybridize. In this case small differences between calculated and experimental frequencies result in large differences in the eigenvectors and the comparison between predicted and calculated phonon intensities is not very meaningful. However, the zone boundary is a high symmetry point where different phonons cannot hybridize because they belong to different symmetry classes. In this case their eigenvectors are fixed by symmetry, so it is possible to use phonon intensities to establish a correspondence between observed and calculated phonons. What we find is that the disagreement between calculated and observed frequencies of $\Sigma_3$-symmetry at the **q**=(0.5, 0.5, 0) zone boundary is stronger than one might guess from inspection of Fig. 1: when considering the eigenvectors, the data points shown at 18 meV and 23 meV correspond to calculated frequencies at 23 meV and 20 meV, respectively, i.e. the phonon frequencies are "flipped".

We also found substantial line broadenings for a number of phonons. For instance, an energy scan at a wavevector **Q** = (2.5,1.5,0) and T = 300 K shows a pronounced broadening for a mode at 18 meV (Fig. 2, top), which, based on its intensity, can be unambiguously assigned to Fe vibrations depicted in the inset of Fig. 2. The line broadening of this branch is maximum at the zone boundary, which becomes a reciprocal lattice point in the low temperature orthorhombic phase. As already mentioned above, its frequency is considerably lower than 23 meV calculated by DFT, There is very little change of mode on cooling from 300K to 190 K but its linewidth shrinks considerably below the tetragonal-to-orthorhombic phase transition at 172 K, (Fig. 2, bottom). These observations indicate a close relationship between the line broadening and the structural instability. However, there is no direct relationship between the elongation patterns of the 18 meV mode and the displacements during the phase transition.

We have also carried out spin-polarized DFT-GGA calculations in the orthorhombic phase of $CaFe_2As_2$. The calculated phonon spectra for non-magnetic/spin-polarized structures are shown as dashed lines Fig. 2upper/lower respectively. It appears that the calculated line widths of phonon modes are larger in the orthorhombic phase because the orthorhombic distortion leads to a splitting of modes. The agreement between our experimental results and the calculation is poor, which further suggest anomalous phonons in CaFe$_2$As$_2$. Unlike the case of BaFe$_2$As$_2$ [11], including magnetism does not improve agreement with experiment (Fig. 2 lower).

It may seem that phonon anomalies are expected at the (0.5 0.5 0) zone boundary point, because it becomes a zone center point in the low temperature deformed orthorhombic structure. However, the mode that corresponds to the lattice deformation at the tetragonal-orthorombic transition is the long wavelength TA phonon (shear mode) along the (0 1 0)-direction. This would be the corresponding soft mode if the phase transition were of second order. However, these phonons do not soften, which is consistent with the first order character of the transition. There is, however, a jump of the slope of this branch at T$_s$ (Fig. 3). The TA 110 branch does show some softening at the zone boundary on approaching the phase transition from above albeit the effect is not very strong (Fig. 3).

Very strong line broadenings were observed not only for the $\Sigma_3$- (TO) modes discussed above, but also for the high-frequency $\Sigma_1$- (LO) modes. Here, the Fe atoms move along 110 towards the As atoms and the As atoms move along 001 towards Fe atoms, which explains the high frequency of these modes. Again, the broadening is maximum at the zone boundary, the full width at half maximum reaching 20 % of the phonon frequency. In contrast to the $\Sigma_3$-phonons, the linewidth of the $\Sigma_1$-modes decreases only slightly on cooling. Our DFT calculations underestimate the linewidth of both the $\Sigma_1$- and the $\Sigma_3$-modes by a huge factor (> 30), which remains to be understood.

The large phonon linewidths in CaFe$_2$As$_2$ could be due to strong anharmonicity or to strong electron-phonon coupling. Anharmonic vibrational behavior appears in very soft compounds, (e.g. potassium), very close to the melting point or close to structural phase transitions involving soft phonons. CaFe$_2$As$_2$ is not soft but it is close to two structural phase transitions, the pressure-induced first order structural phase transition to the nonmagnetic "collapsed" phase and the ambient pressure transition on cooling to the magnetic orthorhombic phase. Since these phase transitions are of first order, this does not necessarily imply a strong anharmonicity; moreover, strong anharmonic effects appear in compounds showing a second order structural phase transformation only for phonons with displacement patterns that correspond to atomic displacements during the phase transition. The displacement patterns of the broadest phonons are not of that sort for either of the two transitions. Thus simple anharmonicity is unlikely to account for the observed linewidths. Strong coupling of phonons to electron-hole excitations is another possibility. However, DFT calculations predict that this coupling should be very weak [9]. Furthermore, no strong phonon broadenings appear in $BaFe_2As_2$ [11], which has similar electronic and crystal structure, and also orders magnetically below 170 K. Also, including magnetic degree of freedom into the calculation for $BaFe_2As_2$ greatly improves the agreement with experimental phonon frequencies. Since the main difference between $BaFe_2As_2$ and $CaFe_2As_2$ that the former is not close to the "collapsed" high pressure phase, this points to the proximity to the "collapsed" high pressure phase as the most probable explanation of the phonon anomalies in $CaFe_2As_2$. A mechanism for this behavior, which accounts for the difference between $CaFe_2As_2$ and $BaFe_2As_2$, has been proposed in [18]. On the other hand, our observation that the broadening of the 18 meV mode becomes much smaller in the magnetic phase leads to the conclusion that the proximity to the magnetic phase is important. Thus more work is necessary to understand the effects we report here. In any case, our findings indicate that the interplay between magnetism and the lattice, which is behind both transitions, is in some way responsible for the anomalous phonons in $CaFe_2As_2$. That is to say, the coupling of the vibrational and the electronic degrees of freedom is stronger than calculated by DFT, and hence phonons might play an important role in superconductivity in the doped compounds.

TABLE I. Comparison between the experimental and calculated structural parameters of the tetragonal phase (*I4/mmm*) of CaFe$_2$As$_2$. The ambient pressure data were taken from Goldmann et al. [7]. The high pressure data were taken from Kreyssig et al. [6].

|  | Experimental P=0, T=300 K [7] | Experimental P=0.63 GPa, T=50 K [6] | DFT-LDA optimized structure |
|---|---|---|---|
| a (Å) | 3.874 | 3.978 | 3.910 |
| c (Å) | 11.664 | 10.607 | 10.264 |
| c/a | 3.011 | 2.676 | 2.625 |
| bond-angle As-Fe-As | 110° | 116° | 119° |
| ,, | 109° | 106° | 105° |

FIG. 1. Comparison of experimentally determined phonon frequencies (solid circles) in the (100), (001) and (110) directions at T = 300 K with results of density functional theory (solid lines). The calculations were based on the experimental crystal structure. The 15 phonon modes along the $\Delta$(100), $\Lambda$(001) and $\Sigma$(110) directions can be classified as $\Delta$: $5\Delta_1 + 2\Delta_2 + 5\Delta_3 + 3\Delta_4$; $\Lambda$: $4\Lambda_1 + \Lambda_2 + 5\Lambda_3$; $\Sigma$: $4\Sigma_1 + 2\Sigma_2 + 4\Sigma_3 + 5\Sigma_4$.

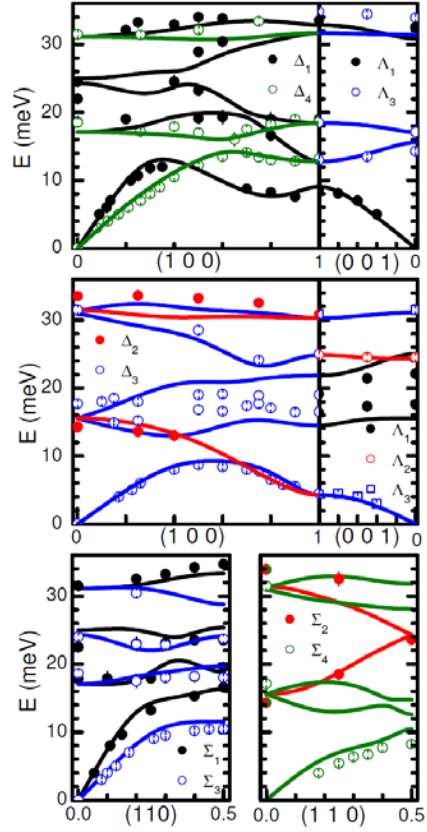

FIG. 2. Energy scans taken at Q = (2.5,1.5,0) at room temperature and at a temperature far below the structural phase transformation. The calculated phonon structure factors for non-magnetic and spin-polarized are shown in upper and lower panels respectively. For better visibility the calculated profiles (dashed lines) in the lower and upper panels are shifted down by 200 counts. The insert in the upper panel shows the q-dependence of the phonon line widths of the branch around 18 meV (red dots) and around 24 meV (blue dots). The lines were obtained by linear regression. The insert in the lower panel shows the displacement pattern of the zone boundary mode at E = 18 meV. Only the Fe atoms are shown. All other atoms are at rest for this mode.

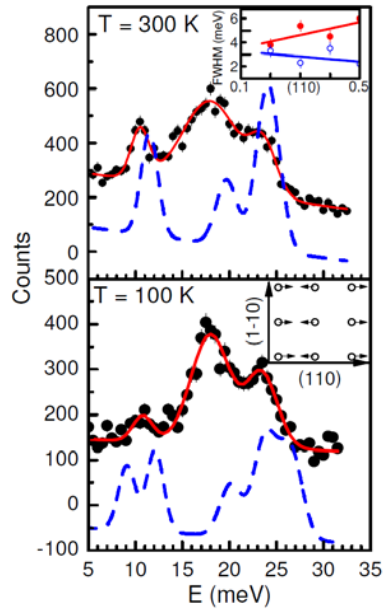

FIG. 3. Temperature dependence of the zone boundary frequency of the TA110 ($\Sigma_3$) (with polarization (1-10) (blue dots, left hand scale) and of the TA100 frequency at q = (0.2,0,0) ($\Delta_3$) (green dots, right hand scale).

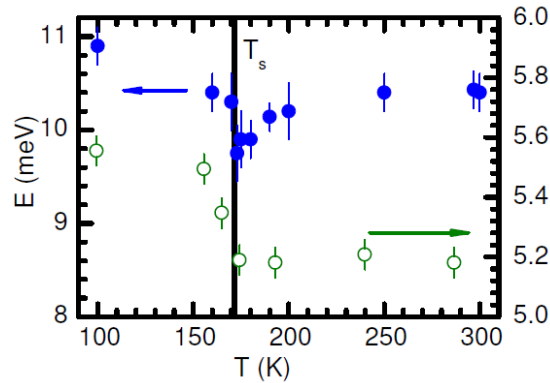